\pdfoutput=1
\PassOptionsToPackage{para}{footmisc}
\documentclass{template/easychair}

\usepackage{doc}
\usepackage{amsmath,amssymb,amsfonts}
\usepackage{todonotes}
\usepackage[font=small,labelfont=bf]{caption}
\usepackage[binary-units,per-mode=symbol]{siunitx}
\usepackage{tikz}
\usetikzlibrary{patterns}
\usepackage{pgfplots}
\pgfplotsset{compat=newest}
\usepgfplotslibrary{units}
\makeatletter
\pgfplotsset{
 unit code/.code 2 args=
   \begingroup
   \protected@edef\x{\endgroup\si{#2}}\x
}
\makeatother
\usetikzlibrary{pgfplots.groupplots}
\definecolor{lightcorered}{RGB}{250,150,150}
\definecolor{corered}{RGB}{200,0,0}
\definecolor{coredarkgray}{RGB}{51,51,51}
\definecolor{coreblue}{RGB}{0, 46, 125}
\definecolor{coregreen}{RGB}{112,173,71}
\usepackage[nohyperlinks,nolist]{acronym} %\ac{} ausgabe, \acl{} langausgabe, \acf{zweites erstes mal}, 
\usepackage{wrapfig}

\newcommand{\q}[1]{``#1''}

\title{
  Simulation-based Evaluation of a Synchronous Transaction Model for Time-Sensitive Software-Defined Networks
}

\author{ 
Tobias Haugg \and Mohammad Fazel Soltani \and Timo H\"ackel \and Philipp Meyer \and\\Franz Korf \and Thomas C. Schmidt
}

\institute{ 
 \href{http://www.haw-hamburg.de/ti-i}{\textit{Dept. Computer Science}},
\href{http://www.haw-hamburg.de/ti-i}{\textit{Hamburg University of Applied Sciences}}, Germany\\
\{\href{mailto:tobias.haugg@haw-hamburg.de}{tobias.haugg},
  \href{mailto:mohammadfazel.soltani@haw-hamburg.de}{mohammadfazel.soltani},
  \href{mailto:timo.haeckel@haw-hamburg.de}{timo.haeckel},
  \href{mailto:philipp.meyer@haw-hamburg.de}{philipp.meyer},\\
  \href{mailto:franz.korf@haw-hamburg.de}{franz.korf},
  \href{mailto:t.schmidt@haw-hamburg.de}{t.schmidt}\}@haw-hamburg.de
}

\authorrunning{Haugg et al.}

\titlerunning{Synchronous Transaction Model for TSSDNs}

\begin{document}

\maketitle

\begin{abstract}
Real-time networks based on Ethernet require robust quality-of-service for time-critical traffic.
The \acf*{TSN} collection of standards enables this in real-time environments like vehicle on-board networks. 
Runtime reconfigurations in TSN must respect the deadlines of real-time traffic.
\acf*{SDN} moves the control plane of network devices to the SDN controller, making these networks programmable.
This allows reconfigurations from a central point in the network.
In this work, we present a transactional model for network reconfigurations that are synchronously executed in all network devices. 
We evaluate its performance in a case study against non-transactional reconfigurations and show that synchronous transactions enable consistency for reconfigurations in TSN without increased latencies for real-time frames. 
\end{abstract}

% The table of contents below is added for your convenience. Please do not use
% the table of contents if you are preparing your paper for publication in the
% EPiC Series or Kalpa Publications series

% \setcounter{tocdepth}{2}
% {\small
% \tableofcontents}

%\section{To mention}
%
%Processing in EasyChair - number of pages.
%
%Examples of how EasyChair processes papers. Caveats (replacement of EC
%class, errors).

%------------------------------------------------------------------------------

%%%% 	ACR			%%%%
% !TEX root = ../main.tex
\begin{acronym}
	% A
	\acro{ACC}[ACC]{Adaptive Cruise Control}
	\acro{ACDC}[ACDC]{Automotive Cyber Defense Center}
	\acro{ACL}[ACL]{Access Control List}
	\acro{ADS}[ADS]{Anomaly Detection System}
	\acroplural{ADS}[ADSs]{Anomaly Detection Systems}
	\acro{ADAS}[ADAS]{Advanced Driver Assistance Systems}
	\acro{API}[API]{Application Programming Interface}
	\acro{AVB}[AVB]{Audio Video Bridging}
	\acro{ARP}[ARP]{Address Resolution Protocol}
	% B
	\acro{BE}[BE]{Best-Effort}
	% C
	\acro{CAN}[CAN]{Controller Area Network}
	\acro{CBM}[CBM]{Credit Based Metering}
	\acro{CBS}[CBS]{Credit Based Shaping}
	\acro{CNC}[CNC]{Central Network Controller}
	\acro{CMI}[CMI]{Class Measurement Interval}
	\acro{CoRE}[CoRE]{Communication over Realtime Ethernet}
	\acro{CT}[CT]{Cross Traffic}
	\acro{CM}[CM]{Communication Matrix}
	% D
	\acro{DoS}[DoS]{Denial of Service}
	\acro{DDoS}[DDoS]{Distributed Denial of Service}
	\acro{DPI}[DPI]{Deep Packet Inspection}
	% E
	\acro{ECU}[ECU]{Electronic Control Unit}
	\acroplural{ECU}[ECUs]{Electronic Control Units}
	% F
	\acro{FDTI}[FDTI]{Fault Detection Time Interval}
	\acro{FHTI}[FHTI]{Fault Handling Time Interval}
	\acro{FRTI}[FRTI]{Fault Reaction Time Interval}
	\acro{FTTI}[FTTI]{Fault Tolerant Time Interval}
	% G
	\acro{GCL}[GCL]{Gate Control List}
	% H
	\acro{HTTP}[HTTP]{Hypertext Transfer Protocol}
	\acro{HMI}[HMI]{Human-Machine Interface}
	\acro{HPC}[HPC]{High-Performance Controller}
	% I
	\acro{IA}[IA]{Industrial Automation}
	\acro{IDS}[IDS]{Intrusion Detection System}
	\acroplural{IDS}[IDSs]{Intrusion Detection Systems}
	\acro{IEEE}[IEEE]{Institute of Electrical and Electronics Engineers}
	\acro{IoT}[IoT]{Internet of Things}
	\acro{IP}[IP]{Internet Protocol}
	\acro{ICT}[ICT]{Information and Communication Technology}
	\acro{IVNg}[IVN]{In-Vehicle Networking}
	\acro{IVN}[IVN]{In-Vehicle Network}
	\acroplural{IVN}[IVNs]{In-Vehicle Networks}
	%J
	% L
	\acro{LIN}[LIN]{Local Interconnect Network}
	% M
	\acro{MOST}[MOST]{Media Oriented System Transport}
	% N
	\acro{NADS}[NADS]{Network Anomaly Detection System}
	\acroplural{NADS}[NADSs]{Network Anomaly Detection Systems}
	% O
	\acro{OEM}[OEM]{Original Equipment Manufacturer}
	\acro{OTA}[OTA]{Over-the-Air}
	%P
	\acro{P4}[P4]{Programming Protocol-independent Packet Processors}
	\acro{PCP}[PCP]{Priority Code Point}
	% R
	\acro{RC}[RC]{Rate-Constrained}
	\acro{REST}[ReST]{Representational State Transfer}
	\acro{RPC}[RPC]{Remote Procedure Call}
	% S
	\acro{SDN}[SDN]{Software-Defined Networking}
	\acro{SDN4CoRE}[SDN4CoRE]{Software-Defined Networking for Communication over Real-Time Ethernet}
	\acro{SIEM}[SIEM]{Security Information and Event Management}
	\acro{SOA}[SOA]{Service-Oriented Architecture}
	\acro{SOC}[SOC]{Security Operation Center}
	\acro{SOME/IP}[SOME/IP]{Scalable service-Oriented MiddlewarE over IP}
	\acro{SR}[SR]{Stream Reservation}
	\acro{SRP}[SRP]{Stream Reservation Protocol}
	\acro{SW}[SW]{Switch}
	\acroplural{SW}[SW]{Switches}
	% T
	\acro{TAS}[TAS]{Time-Aware Shaping}
	\acro{TCP}[TCP]{Transmission Control Protocol}
	\acro{TDMA}[TDMA]{Time Division Multiple Access}
	\acro{TSN}[TSN]{Time-Sensitive Networking}
	\acro{TSSDN}[TSSDN]{Time-Sensitive Software-Defined Networking}
	\acro{TT}[TT]{Time-Triggered}
	\acro{TTE}[TTE]{Time-Triggered Ethernet}
	% U
	\acro{UDP}[UDP]{User Datagram Protocol}
	\acro{UN}[UN]{United Nations}
	% Q
	\acro{QoS}[QoS]{Quality-of-Service}
	% V
	\acro{V2X}[V2X]{Vehicle-to-X}
	%W
	\acro{WS}[WS]{Web Services}
	% Z
	\acro{ZC}[ZC]{Zone Controller}

\end{acronym}

%!TEX root = ../main.tex

\section{Introduction}%
\label{sec:introduction}
In recent years, Ethernet is playing an increasingly important role, even finding its way into real-time applications.
Complementary protocols such as IEEE~802.1Q~\cite{ieee8021q-18} \ac{TSN} have proven to meet the real-time and robustness requirements of these environments.
On the other hand, \ac{SDN}~\cite{mabpp-oeicn-08} has revolutionized campus and data center networks.
Separating the control from the data plane of network devices at a central control unit enables simple and fast-forwarding at devices, while high level control applications with global network knowledge can steer the entire network.

One challenge in frequently changing networks is reconfiguration at runtime, which is often necessary when flows are added or removed sporadically, and down time is not an option.
The biggest problem faced is maintaining a consistent state without affecting network performance~\cite{TBFLCDRSDN18}, which is especially important in real-time environments.
Transactions can solve these problems in distributed systems by providing the \textit{ACID} properties \textit{atomicity}, \textit{consistency}, \textit{isolation}, and \textit{durability}~\cite{tanenbaumDS17}.
This way transactions guarantee that the network is always in a consistent state no matter whether a reconfiguration succeeds or not.
In real-time environments the synchronous execution of transactions over multiple network devices also plays an important role when configuring traffic with low-latency requirements and no tolerance to packet loss.

In this work, we present our time synchronous transaction model and compare it with a non-transactional reconfiguration approach, both of which we implement using NETCONF and study their performance.
To evaluate the different implementations, we use the discrete event simulation platform OMNeT++.
Our implementations extend our previously published simulation framework \textit{\acf{SDN4CoRE}}~\cite{hmks-smsdn-19}, which enables the simulation of programmable (software-defined) real-time Ether-networks. 
% It uses the OpenFlow framework and provides additional programming mechanisms such as NETCONF to enable programming of real-time Ethernet components via controller applications.

The paper is structured as follows. 
Section~\ref{sec:background_related_work} provides background knowledge and related work. 
Section~\ref{sec:concept} shows our concept of synchronous transactions.
Section~\ref{sec:eval} compares the performance of synchronous transactions with non-transactional network reconfiguration in a case study.
Finally, section~\ref{sec:conclusion_outlook} draws a conclusion and gives an outlook on future work.

%!TEX root = ../main.tex

\section{Background \& Related Work}%
\label{sec:background_related_work}
Our time-synchronous transaction model enables reconfiguration of networks combining \ac{TSN} and \ac{SDN}.
This section introduces into \ac{TSN}, \ac{SDN}, Transactions, and related work.

\subsection{Time-Sensitive Networking}
\label{subsec:background_tsn}
The \ac{TSN} real-time Ethernet protocol is a collection of standards specifically designed for use in industrial control equipment and communication networks within vehicles.
\ac{TSN} is defined in IEEE~802.1Q~\cite{ieee8021q-18} and bases on Ethernet frames with Q-Tag that have a VLAN ID and priority code point (from 0 - low to 7 - high).
To enable heterogenous \ac{QoS} guarantees for each individual priority each port has a frame selector instance that decides when which packet is sent.

\begin{wrapfigure}{R}{0.65\linewidth}
    \centering
    \includegraphics[width=1\linewidth, trim= 0.62cm 0.62cm 0.62cm 0.62cm, clip=true]{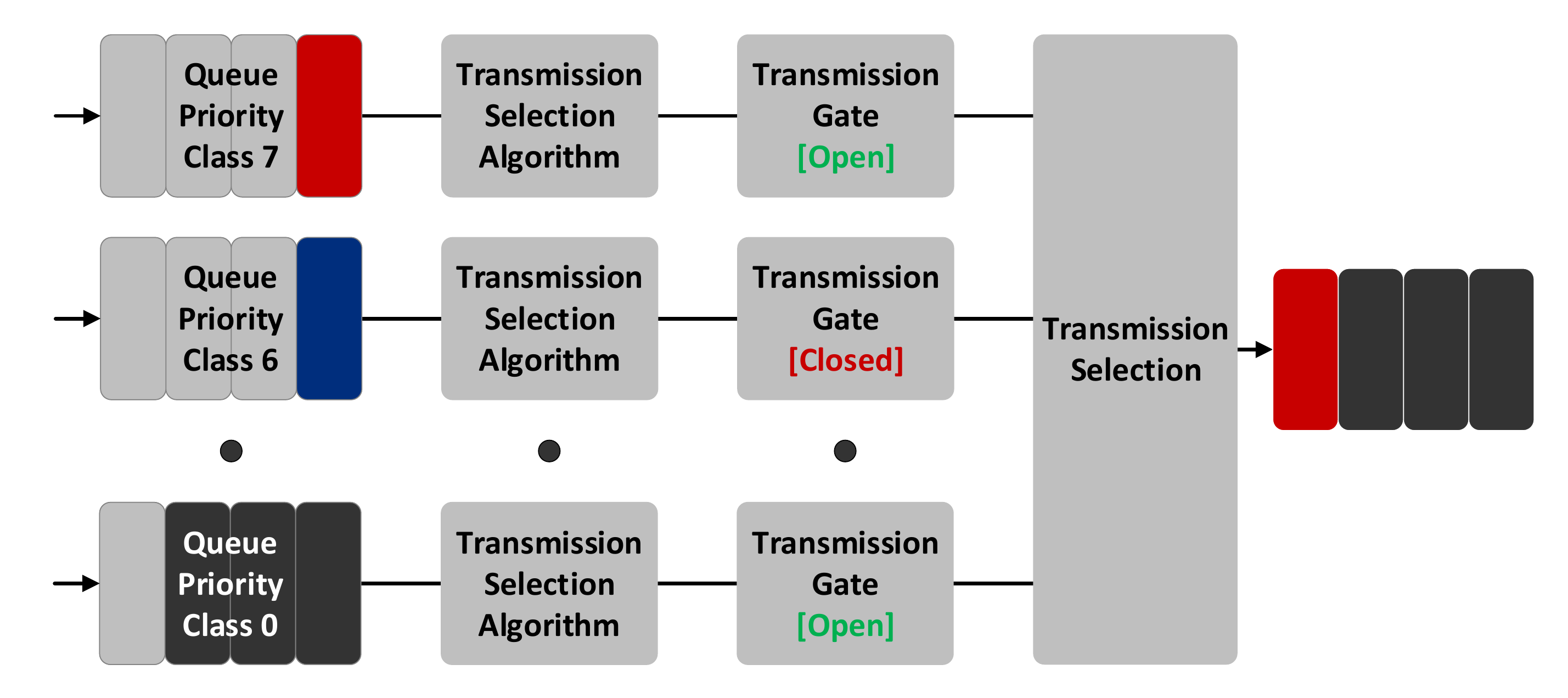}
    \caption{\ac{TSN} scheduling and forwarding \cite{ieee8021q-18}}
    \label{fig:tsn_transmission}
\end{wrapfigure}

Figure~\ref{fig:tsn_transmission} shows the frame selection. Frames can be assigned to one of eight priorities.
For each priority, the frame selector has a queue, a Transmission Selection Algorithm, and a Transmission Gate.
If more than one path is open to the Transmission Selection, the frame is taken from the queue with the highest priority and sent to the interface.

Transmission Selection Algorithm instances can implement different frame selection algorithms.
An example of this is the \ac{CBS} algorithm, which ensures compliance with a reserved bandwidth.
If the Transmission Selection Algorithm allows a frame to be sent, the next thing to consider is the transmission gate.

The gates have one of two possible states \q{OPEN} or \q{CLOSED}.
Sending a frame is only allowed when the \q{OPEN} state is active.
The state change can be timed based on a configuration called \ac{GCL}.
This enables scheduling of \ac{TDMA} traffic by opening individual Transmission Gates in specific time intervals.
The accuracy of such a schedule depends on which devices are involved in the scheduling. 
In a fully scheduled network, all endpoints and switches implement a common \ac{TDMA} schedule.

\subsection{Software-Defined Networking}
\acf{SDN}~\cite{mabpp-oeicn-08} separates the control logic (control plane) from the underlying switches that forward the traffic (data plane)~\cite{krvra-sdncs-15}.
Network devices become programmable by a central \ac{SDN} controller using open standard protocols such as OpenFlow.
OpenFlow switches forward incoming packets based on a programmable flow table.
A flow entry matches a subset of Layer~2 to Layer~4 header fields and contains actions, such as discard or forward, that are performed when traffic matches.
\ac{SDN} controller applications implement the behavior of the network, e.g. routing protocols, and populate the flow tables in the switches.

\ac{SDN} can reduce the complexity and increase the adaptability of real-time Ether-networks to cope with the growing dynamics introduced, for example, by \ac{SOA}.
In previous work, we integrated \ac{TSN} with \ac{SDN}~\cite{hmks-snsti-19} to control dynamic bandwidth reservation with a central \ac{SDN} controller without delay penalty for real-time communication. 
Dynamic control of scheduled traffic is more complex because it must be coordinated between multiple senders across multiple hops in the network.
In this paper, we explore different solutions for network schedule reconfiguration.

\subsection{Transactional Network (Re-)Configuration}
Transactions are commonly used to coordinate changes in distributed systems~\cite{tanenbaumDS17}.
The ACID properties can guarantee the validity of a transaction despite the occurrence of errors:
\begin{itemize}
    \item \textit{Atomicity}: A transaction composed of multiple operations either succeeds completely, or fails completely. If a transaction fails the system is left unchanged.
    \item \textit{Consistency}: A transaction always transfers the system from one valid state to another.
    \item \textit{Isolation}: Concurrent transactions behave exactly as if they were executed sequentially.
    \item \textit{Durability}: When a transaction has been committed is remains committed until the next changes are made to the system.
\end{itemize}

The properties of ACID transactions can be used in SDN environments to maintain the consistency of the network, e.g. by extending the OpenFlow protocol~\cite{transactional_updates_in_sdn}.
Cui et al. show how transactions can bundle changes to multiple flow rules of a network function~\cite{TBFLCDRSDN18}.
OpenFlow is well suited to update flows in the network.
It does not support TSN \ac{GCL} configuration. 

The NETCONF protocol (RFC~6241~\cite{RFC-6241}) can be used for reconfiguration of \ac{TSN} modules and the OpenFlow table at the same time via existing YANG data models.
NETCONF provides useful features to implement transactions.
Each device has multiple configuration data stores.
Each data store can be locked by exactly one client guaranteeing isolation until the store is unlocked.
One data store contains the active device configuration (\textit{running}) and changes to it are directly applied and activated.
\textit{Candidate} datastores contain a copy of the running configuration and changes are applied only to the copy so that the full configuration can be verified and later committed and thus applied to the running datastore.

In this work, we extend the NETCONF transaction mechanism to allow synchronously executed transactions over multiple network devices.

\section{Synchronous Transactions for Network Reconfiguration}%
\label{sec:concept}

Our time-synchronous transaction model is coordinated by an \ac{SDN} controller that reconfigures a set of switches using the NETCONF protocol.
\newline\newline
\begin{wrapfigure}{R}{0.5\linewidth}
    \centering
    \includegraphics[width=1\linewidth, trim= 0.6cm 0.6cm 0.6cm 0.6cm, clip=true]{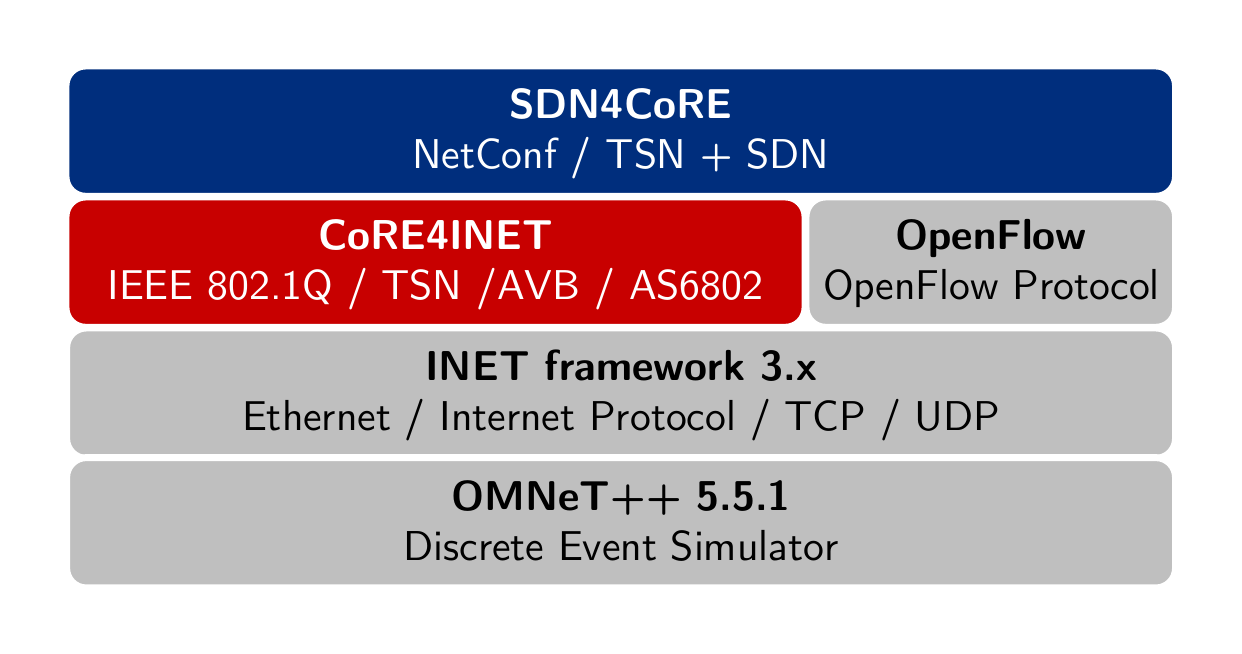}
    \caption{Simulation environment}
    \label{fig:sim_env}
    \vspace{-10pt}
\end{wrapfigure}

Figure~\ref{fig:sim_env} shows the simulation environment used for our implementation, which is based on the discrete event simulator OMNeT++\footnote{\url{omnetpp.org}} (version 5.5.1).
The INET framework\footnote{\url{inet.omnetpp.org}} (version 3.x) implements internet and transport protocols, our fork of the OpenFlowOmnetSuite\cite{kj-oeojr-13} enables \ac{SDN} with the OpenFlow protocol, our CoRE4INET framework~\cite{mkss-smcin-19} implements TSN. 
We incorporated the transaction model into the SDN4CoRE framework, which we presented in previous work~\cite{hmks-smsdn-19}.
It adds a programming option for real-time Ether-networks and provides an implementation of the NETCONF protocol.
Our frameworks are published and available as open source\footnote{\url{sim.core-rg.de} and \url{github.com/CoRE-RG}}.

A few assumptions were made to simplify the implementation: 
A reliable transport protocol is used, there are no hardware or network errors, and each switch supports the same TSN and SDN features.
The clocks of all devices in the network are synchronized with a maximum inaccuracy of $\pm$\SI{500}{\nano\second}.  
%Furthermore, any error after the commit execution, as well as hardware or network errors will not be handled by the transaction model.
% To handle a crash caused from the tcp protocol the transaction model ether tries a reconnect or reverts the entire transaction.

\begin{figure}[ht]
    \centering
    \includegraphics[width=1.0\linewidth, trim= 1.4cm 14.6cm 1.1cm 0.9cm, clip=true]{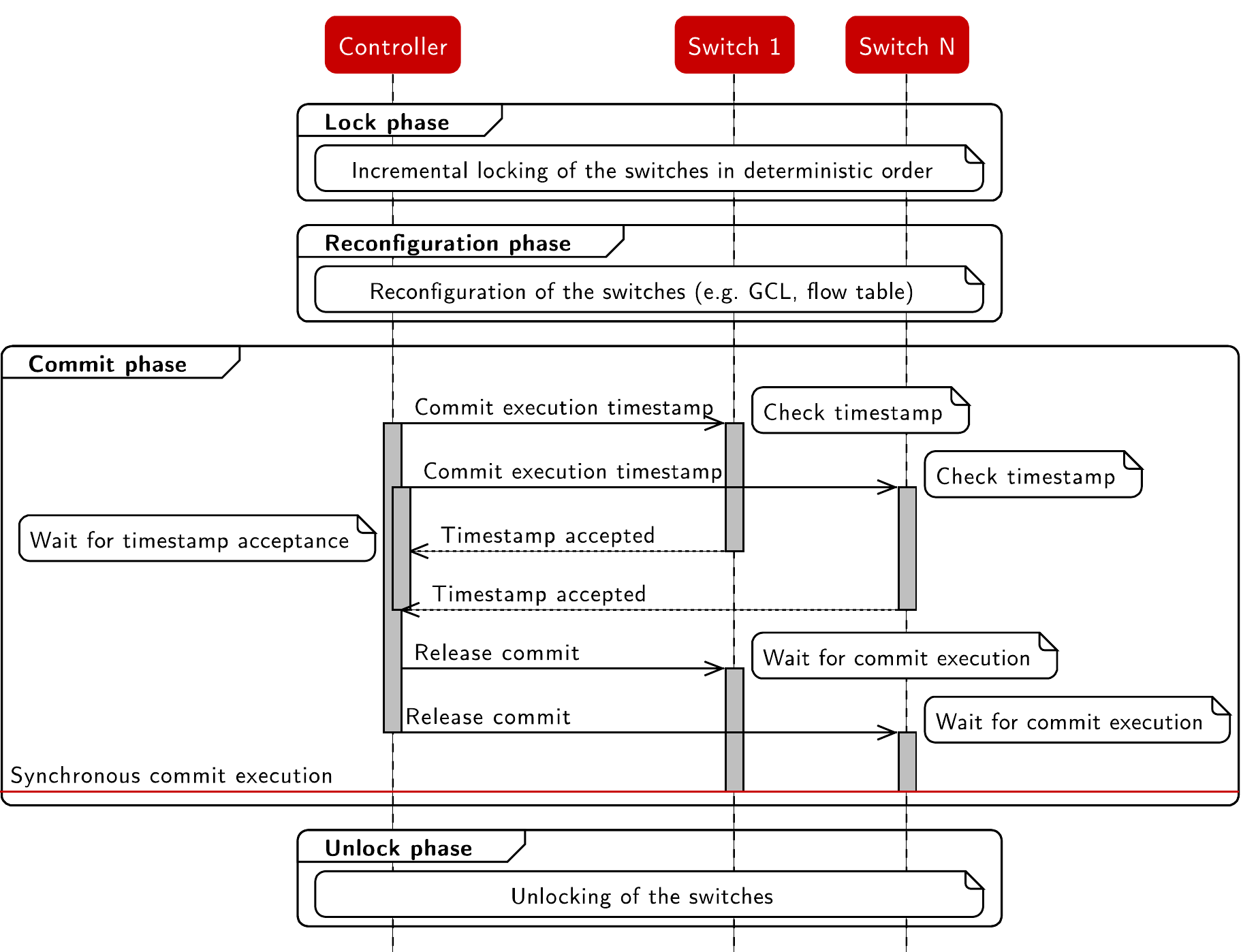}
    \caption{Simplified sequence diagram of the four phases in a successful transaction. }
    \label{fig:synctrans_sequence}
\end{figure}

Our synchronous transaction model consists of four phases: \textit{Lock}, \textit{reconfiguration}, \textit{commit}, and \textit{unlock}. 
Figure \ref{fig:synctrans_sequence} depicts a successful transaction in a sequence diagram.
The lock, reconfiguration, and unlock phases use standard NETCONF operations and are thus shown in simplified form.
The commit phase is shown in detail.
It implements the two phase commit protocol~\cite{tanenbaumDS17} and adds support for synchronized reconfiguration.

At first, the lock phase reserves all resources needed for the transaction. 
The controller attempts to acquire a lock for each switch whose configuration is supposed to be modified. 
To avoid deadlocks, the switches are contacted one after the other in the order of their MAC addresses. 
If a lock cannot be acquired, all previously obtained locks are released, and the transaction terminates unsuccessfully.

The second phase is the reconfiguration phase. 
For each switch, a candidate configuration is instantiated that is a copy of the previously locked running configuration. 
The candidate configuration is locked immediately after it is instantiated and is used to apply any reconfigurations.
If an error occurs during the copy operation, while locking the candidate configuration, or during any of the reconfigurations, all the changes made are undone and the running configuration remains unchanged.

In the third phase, called commit phase, the candidate configuration of each switch becomes the new running configuration. 
A network-wide commit execution time is used to synchronously activate the new configurations in all switches. 
The controller calculates the commit execution time and performs a two-stage commit, first checking that each switch is ready and accepts the timestamp, and then releasing the commit for execution at the agreed time.
% The controller sends the commit timestamp to all the switches involved. 
% When the switches accept the timestamp, the controller sends a commit release message to them indicating that the configuration should be activated at the agreed timestamp.
The timestamp for the commit execution is calculated based on the Worst-Case Execution Time (WCET) of the commit phase.
For this, the accumulated maximum latencies for all messages and processing times at the controller and switches are summed up.  
The WCET is independent from the complexity of the reconfigurations since they have already been applied to the candidate configuration in the previous phase. 
This time span can then be used to either directly determine a commit execution timestamp or the next possible start of a network period for the commit execution.
All devices apply updates in the candidate configuration to the running configuration when the synchronous timestamp is reached.

At the end of the transaction, the fourth phase unlocks all resources and cleans up after the transaction. 
The unlock phase is always executed, regardless of whether any of the previously described phases failed. 
All unconfirmed changes and the old configuration are deleted. 
At last, all previously acquired locks are released.

%!TEX root = ../main.tex
\section{
    % Case Study: (In-)Consistent Network Reconfiguration
    Case Study on Network Reconfiguration Consistency
    }%
\label{sec:eval}
We compare the performance of the transactional model with a non-transactional approach using the impact of reconfigurations on the end-to-end latency of frames from real-time flows.
% We compare the performance of the transactional model with the non-transactional approach.
% The impact of reconfigurations is compared using the end-to-end latency of frames belonging to real-time flows.

\begin{wrapfigure}{R}{0.55\linewidth}
    \centering
    \includegraphics[width=1\linewidth]{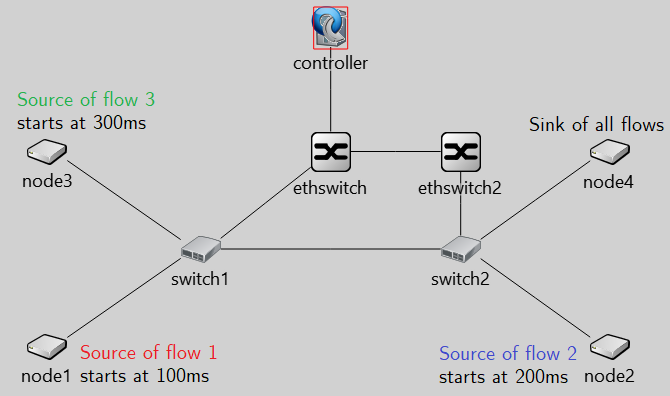}
    \caption{Network topology of the case study}
    \label{fig:eval_topology}
\end{wrapfigure}
\begin{figure}[ht]
    \centering
    \begin{tikzpicture}
        \begin{axis}[
            clip mode = individual,
            width = \linewidth ,
            height = 0.37\linewidth ,
                change y base,
                y SI prefix=milli,
                y unit=\milli\second,
                x unit=\second,
                xlabel={Simulation time},
                ylabel={End-to-end latency},
                %legend pos=north east,
                legend style={at={(axis cs:0.49,0.0008)},anchor=south east},
                legend cell align={left},
                ymax=0.0035,
                xmin=0.051,
                xmax=0.499,
                xtick distance=0.05
                ]
                %latencies
                \addplot [
                    ultra thick, 
                    corered
                    ] table[x,y, col sep = comma] {data/NonTransactionalAndNonTimesynchronous/latency_realtimeflow1_node1_to_node3.csv};
                    \addplot [
                        % mark=*, 
                        very thick,
                        coreblue
                        ] table[x,y, col sep = comma] {data/NonTransactionalAndNonTimesynchronous/latency_realtimeflow2_node2_to_node3.csv};
                        \addplot [
                            very thick, 
                            coregreen
                            ] table[x,y, col sep = comma] {data/NonTransactionalAndNonTimesynchronous/latency_realtimeflow3_node4_to_node3.csv};                
                            %changes
                            %A1 -> Switch 1 Flowentry in Flowtable
                            % \addplot [cyan, thin, update limits=false] coordinates {(0.102011799,0) (0.102011799,0.0045)};
                            %A2 -> Switch 1 modify Gate Control List
                            % \addplot [cyan, thin, update limits=false] coordinates {(0.104011799,0) (0.104011799,0.0045)};
                            %A1 -> Switch 2 Flowentry in Flowtable
                            \addplot [gray, thick, update limits=false] coordinates {(0.102018762,0) (0.102018762,0.0045)};
                            %A2 -> Switch 2 modify Gate Control List
                            \addplot [gray, thick, update limits=false] coordinates {(0.104018762,0) (0.104018762,0.0045)};
                            \node [anchor=north, clip=false] at (axis cs:0.102018762,0.0039) {U1,U2};
                            %A3 -> Switch 2 Flowentry in Flowtable
                            \addplot [gray, thick, update limits=false] coordinates {(0.202016794,0) (0.202016794,0.0045)};
                            %A4 -> Switch 2 modify Gate Control List
                            \addplot [gray, thick, update limits=false] coordinates {(0.204016794,0) (0.204016794,0.0045)};
                            \node [anchor=north, clip=false] at (axis cs:0.202016794,0.0039) {U3,U4};
                            %A5 -> Switch 1 Flowentry in Flowtable
                            % \addplot [cyan, thin, update limits=false] coordinates {(0.302011799,0) (0.302011799,0.0045)};
                            %A6 -> Switch 1 modify Gate Control List
                            % \addplot [cyan, thin, update limits=false] coordinates {(0.304011799,0) (0.304011799,0.0045)};
                            %A5 -> Switch 2 Flowentry in Flowtable
                            \addplot [gray, thick, update limits=false] coordinates {(0.302018762,0) (0.302018762,0.0045)};
                            %A6 -> Switch 2 modify Gate Control List
                            \addplot [gray, thick, update limits=false] coordinates {(0.304018762,0) (0.304018762,0.0045)};
                            \node [anchor=north, clip=false] at (axis cs:0.302018762,0.0039) {U5,U6};
                            \legend{Real-Time Flow 1, Real-Time Flow 2, Real-Time Flow 3}
                        \end{axis}
            \vspace{-5pt}
        \end{tikzpicture}
        \caption{End-to-end latency for non-synchronous, non-transactional network updates}
    \label{fig:latencies_non_sync}
\end{figure}
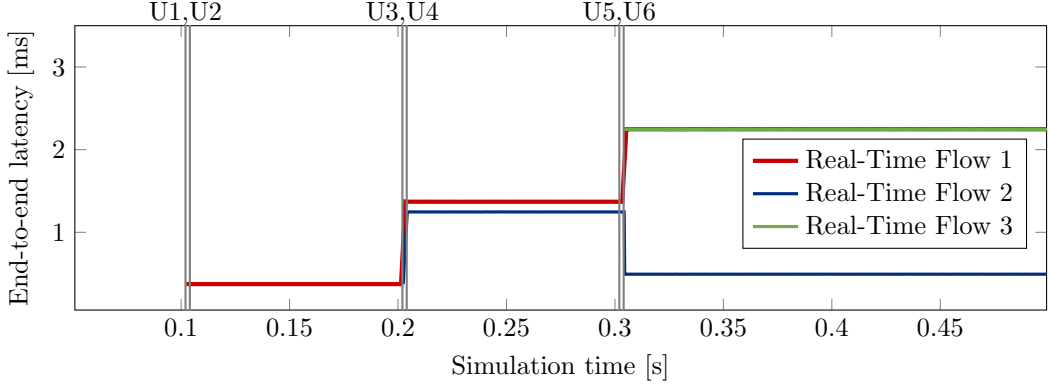
Figure~\ref{fig:eval_topology} shows our evaluation network, which consists of four nodes connected via two \acf{TSSDN} switches.
At the control plane, these switches are connected to a \acf{TSSDN} controller via two Ethernet switches that introduce a separate delay between controller and \acf{TSSDN} switches.
Each link is 10 meters long and has a speed of \SI{100}{\mega\bit\per\second}.
Also, each switch has a forwarding delay of \SI{3}{\micro\second}.% and can hold up to three packets per priority output queue.

Node1, node2, and node3 send one full-size Ethernet packet every millisecond over their respective real-time flows 1, 2, and 3, which are received by node4.
On all devices a \ac{TDMA} schedule is used repeating every \SI{1}{\milli\second}. 
All packets from flow 1, 2 and 3 share the same priority, and therefore the same time slot in the schedule.
The time slot always has to adapt to the number of active real time flows.
Switch1 has to offer a time slot long enough to forward the packets from flow 1 and 3 and switch2 to has to offer enough time to forward flow 1, 2 and 3 during one period. 
To avoid interference from cross traffic, a guard band is added to close all gates and guarantee that the band is clear when real-time traffic arrives.

We calculate the expected latencies of the real-time flows based on the transmission delay over the links, and the forwarding delay in the switches that a packet must pass.
This results in an expected latency of \SI{373.35}{\micro\second} for packets originating from node1 and node3, and \SI{247.9}{\micro\second} for packets from node2.

Traffic at the nodes is started sequentially with an offset of \SI{100}{\milli\second} to observe the effects of the reconfigurations on existing and added traffic. 
After a node is activated, two reconfigurations are performed on the switches.
First, the flow table is modified to forward the new packets correctly; second, the \ac{GCL} schedule is updated to accommodate an additional packet of the new real-time flow in the time slot for scheduled traffic.
A transaction bundles both modifications per activated node, while in the non-transactional approach these modifications are performed independently.
In total, we have six reconfigurations \mbox{(U1-U6)} using the non-transactional approach and three transactions \mbox{(T1-T3)} combining related modifications.

Figure~\ref{fig:latencies_non_sync} shows the end-to-end latency for the non-transactional reconfiguration.
The latency increases drastically during the third and fourth reconfigurations (U3 and U4) and then remains constant until the next flow is added.
During U5 and U6, the latency of flow~1 increases again and then remains constant at a fairly high level together with flow~3.
The latency of flow~2 decreases but is still above the previously calculated latency of \SI{247.9}{\micro\second}.

\begin{figure}[t]
    \centering
    \begin{tikzpicture}
        \begin{axis}[
            clip mode = individual,
            width = \linewidth ,
            height = 0.37\linewidth ,
            change y base,
            y SI prefix=milli,
            y unit=\milli\second,
            x unit=\second,
            xlabel={Simulation time},
            ylabel={End-to-end latency},
            %legend pos=south east,
            legend style={at={(axis cs:0.2107,0.00075)},anchor=east},
            legend cell align={left},
            ymin=0,
            xmin=0.199,
            xmax=0.211,
            xtick distance = 0.002,
            x tick label style={/pgf/number format/fixed relative,/pgf/number format/precision=5},
            ]
            %latencies
        \addplot [
            mark=o, 
            very thick,
            corered
            ] table[x,y, col sep = comma] {data/NonTransactionalAndNonTimesynchronous/latency_realtimeflow1_node1_to_node3.csv};
            \addplot [
                mark=*, 
                very thick,
                coreblue
                ] table[x,y, col sep = comma] {data/NonTransactionalAndNonTimesynchronous/latency_realtimeflow2_node2_to_node3.csv};
                %changes
                %A3 -> Switch 2 Flowentry in Flowtable
                \addplot [gray, thick, update limits=false] coordinates {(0.202016794,0) (0.202016794,0.002)};
                \node [anchor=north, clip=false] at (axis cs:0.202016794,0.00168) {U3};
                %A4 -> Switch 2 modify Gate Control List
                \addplot [gray, thick, update limits=false] coordinates {(0.204016794,0) (0.204016794,0.002)};
                \node [anchor=north, clip=false] at (axis cs:0.204016794,0.00168) {U4};
                \legend{Real-Time Flow 1, Real-Time Flow 2}
        \end{axis}
    \end{tikzpicture}
    \vspace{-5pt}
    \caption{Detailed view of the end-to-end latency during reconfigurations U3 and U4 with non-synchronous, non-transactional network updates}
    \label{fig:latencies_non_sync_detail}
\end{figure}
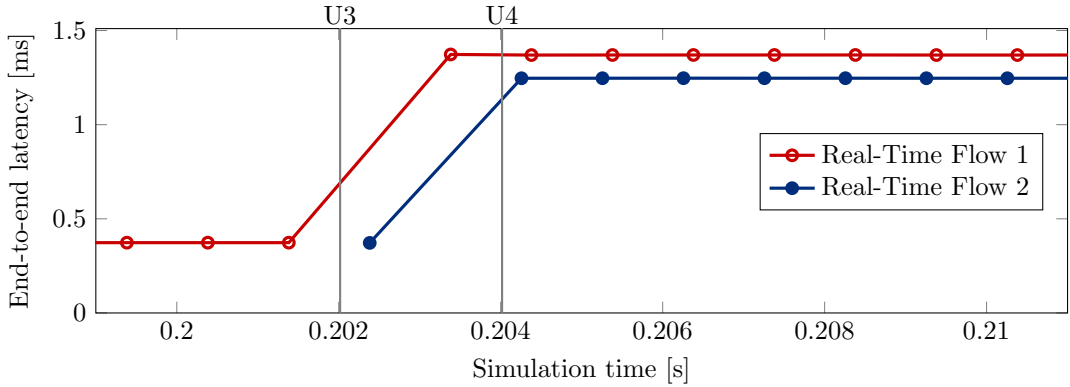
        
Figure~\ref{fig:latencies_non_sync_detail} gives a more detailed view of reconfigurations U3 and U4 between \SI{200}{\milli\second} and \SI{210}{\milli\second} simulation time.
After reconfiguration U3 adds a flow rule for flow~2, the flow is forwarded through switch2.
The latency for flow~1 spikes because the first packet from flow~2 takes away the time slot originally reserved for flow~1 and the \ac{GCL} schedule has not yet been changed by reconfiguration U4. 
Thus, the packet from flow~1 misses its time slot and must wait one period before it can be transmitted.
In the next period, the packet from flow~1 is transmitted, while the newly arrived packet from flow~2 must now wait for one period.
Reconfiguration U4 updates the \ac{GCL} to now accommodate two full-size packets per period, one for flow~1 and one for flow~2. 
This results in constant but severely delayed latencies.
Similar behavior can be observed after U5 and U6.
The drastically increased latencies of flow~1 and 3 show that the time-guarantees of the TSN network are no longer met, making this non-transactional approach inapplicable in real-time networks.
        
\begin{figure}[t]
    \centering
    \begin{tikzpicture}
        \begin{axis}[
            clip mode = individual,
            width = \linewidth ,
            height = 0.37\linewidth ,
            change y base,
            y SI prefix=milli,
            y unit=\milli\second,
            x unit=\second,
            xlabel={Simulation time},
            ylabel={End-to-end latency},
            %legend pos=north east,
            legend style={at={(axis cs:0.49,0.00027)},anchor=south east},
            legend cell align={left},
            ymax=0.00045,
            xmin=0.051,
            xmax=0.499
            ]
            
            %latencies
            \addplot [
                ultra thick,
                corered
                ] table[x,y, col sep = comma] {data/SoP_TransactionalAndTimesynchronous/latency_realtimeflow1_node1_to_node3.csv};
                \addplot [
                    % mark=*, 
                    very thick,
                    coreblue
                    ] table[x,y, col sep = comma] {data/SoP_TransactionalAndTimesynchronous/latency_realtimeflow2_node2_to_node3.csv};
                    \addplot [
                        very thick,
                        coregreen
                        ] table[x,y, col sep = comma] {data/SoP_TransactionalAndTimesynchronous/latency_realtimeflow3_node4_to_node3.csv};                
                        
                        %transactions
                        %TA1 -> Switch 1
                        % \addplot [cyan, thick, update limits=false] coordinates {(0.143999721253,0) (0.143999721253,0.00055)};
                        %TA1 -> Switch 2
                        \addplot [gray, thick, update limits=false] coordinates {(0.144000400966,0) (0.144000400966,0.00055)};
                        \node [anchor=north, clip=false] at (axis cs:0.144000400966,0.000475) {T1};
                        %TA2 -> Switch 2
                        \addplot [gray, thick, update limits=false] coordinates {(0.238000259438,0) (0.238000259438,0.00055)};
                        \node [anchor=north, clip=false] at (axis cs:0.238000259438,0.000475) {T2};
                        %TA3 -> Switch 1
                        % \addplot [cyan, update limits=false] coordinates {(0.34399974328,0) (0.34399974328,0.00055)};
                        %TA3 -> Switch 2
                        \addplot [gray, thick, update limits=false] coordinates {(0.344000238588,0) (0.344000238588,0.00055)};
                        \node [anchor=north, clip=false] at (axis cs:0.344000238588,0.000475) {T3};
                        \legend{Real-Time Flow 1, Real-Time Flow 2, Real-Time Flow 3}          
                    \end{axis} 
                \end{tikzpicture}
\vspace{-5pt}
\caption{End-to-end latency for time-synchronous, transactional network updates}
\label{fig:latencies_sync}
\end{figure}
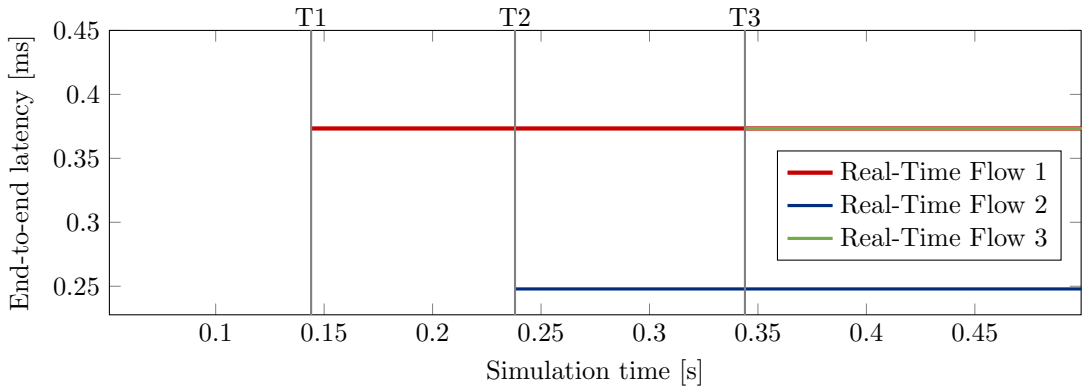
    
Figure~\ref{fig:latencies_sync} shows the results for the time-synchronous transactions.
In the transactional approach, the two reconfigurations that need to be performed for each new node are combined into one transaction. 
In addition, the transaction is executed synchronously on both switches, reducing the period of inconsistency in the network to the accuracy of time synchronization.
With time-synchronous transactions, the latency of all real-time flows stays constant and matches the calculated latencies throughout the simulation.
The difference in latencies compared to the non-transactional approach is caused by the way the reconfigurations are bundled and executed.

Comparing the results of Figure~\ref{fig:latencies_sync} with the non-transactional approach (Figure~\ref{fig:latencies_non_sync}), we can see that they differ not only in the end-to-end latency of the traffic, but also in the execution time of reconfigurations.
Although the reconfiguration starts at the same time in both approaches, we can observe a delay in the execution of the transactions.
This delay is caused by the synchronous commit execution based on the WCET.
% Since a transaction first applies all changes to a copy of the running configuration and then activates all changes at the same time, no inconsistencies can occur due to the activation of the modified copy (candidate configuration).

Our results show that the synchronous transaction model is indeed able to perform network reconfiguration for real-time traffic without delay penalties for individual frames, making it possible to reconfigure TSN networks during runtime.
%!TEX root = ../main.tex
\section{Conclusion \& Outlook}
\label{sec:conclusion_outlook}
In this paper, we presented our approach for synchronous transactional runtime reconfigurations in \ac{TSN}.
Our case study showed, due to inconsistencies in reconfigurations the non-transactional approach led to drastically increased latencies and has been shown to be unsuitable for time critical traffic.
In contrast, the time-synchronous transactional model was able to reconfigure the \ac{TSN} devices during runtime without affecting the latency of existing and added real-time traffic.
By merging related reconfigurations into one transaction that is executed synchronously in all network devices, the network can be seamlessly transition between consistent states.

Future work will investigate the performance of synchronous transactions in \ac{TSN} under various forms of modifications such as deleting and rerouting existing real-time flows.%, as well as decreasing the time slot in a \acf{GCL}.
Furthermore, we will compare several forms of commit synchronization, e.g., commit execution in the middle of a period as opposed to the beginning of the period might be of interest.

%------------------------------------------------------------------------------
\section*{Acknowledgments}
\label{sect:acks}
This work is funded by the Federal Ministry of Education and Research of Germany (BMBF) within the SecVI project.

\label{sect:bib}
\bibliographystyle{plain}
\bibliography{rfcs,HTML-Export/all_generated,bib/special}

%\bibliographystyle{alpha}
%\bibliographystyle{unsrt}
%\bibliographystyle{abbrv}

%------------------------------------------------------------------------------
% Index
%\printindex

%------------------------------------------------------------------------------
\end{document}